# Optimizing Memory Efficiency for Deep Convolutional Neural Networks on GPUs

Chao Li[#] Yi Yang* Min Feng* Srimat Chakradhar* Huiyang Zhou[#]

[#] Department of Electrical and Computer Engineering, North Carolina State University

* Department of Integrated System, NEC Labs America.

[#]{cli17, hzhou} @ ncsu.edu; *{yyang, mfeng, chak} @ nec-labs.com

***Abstract*— Leveraging large data sets, deep Convolutional Neural Networks (CNNs) achieve state-of-the-art recognition accuracy. Due to the substantial compute and memory operations, however, they require significant execution time. The massive parallel computing capability of GPUs make them as one of the ideal platforms to accelerate CNNs and a number of GPU-based CNN libraries have been developed. While existing works mainly focus on the computational efficiency of CNNs, the memory efficiency of CNNs have been largely overlooked. Yet CNNs have intricate data structures and their memory behavior can have significant impact on the performance. In this work, we study the memory efficiency of various CNN layers and reveal the performance implication from both data layouts and memory access patterns. Experiments show the universal effect of our proposed optimizations on both single layers and various networks, with up to 27.9x for a single layer and up to 5.6x on the whole networks.**

## I. Introduction

The success of the deep Convolutional Neural Network (CNN), Alex-Net [12], in the 2012 ImageNet recognition competition has made it as one of the most promising machine learning techniques. In the past few years, many deep neural networks have been developed and the latest CNN powered image recognition even outperformed human vision [10]. There are two main reasons for the success of deep CNNs. The first is large-scale training data sets and the second is large and deep neural network structures. Both require substantial computational and memory throughput. As a result, many-core processors like GPUs, featuring high computational throughput and memory access bandwidth, have become a popular accelerator for deep CNNs. Recently, a number of GPU-based accelerated CNN libraries have been developed. Cuda-convenet [15] was the first highly optimized CNN implementation on GPUs. After that, a number of popular machine learning frameworks such as Torch [7], Theano [1], Caffe [11] have released their own GPU libraries for CNNs. Among them, Caffe is the most popular deep learning framework and has been widely used in the machine learning community. The GPU hardware vendor, NVIDIA, also develops a new library, cuDNN [4], which provides highly optimized and portable GPU kernel functions used in CNNs. Apart from them, there are also recent studies on accelerating CNNs by reducing the arithmetic complexity in convolutional layers [8][16][23], and using coarse-grain parallelism in thread mapping [9]. These existing works mostly focus on the computational efficiency of the network, especially that of convolutional layers. The memory efficiency of the network, however, has been largely overlooked. As deep neural networks have intricate data structure, the performance implication of memory behavior is not straightforward. Our study unveils that there are two aspects that have not been addressed yet pose non-trivial impact on memory efficiency and the overall CNN performance.

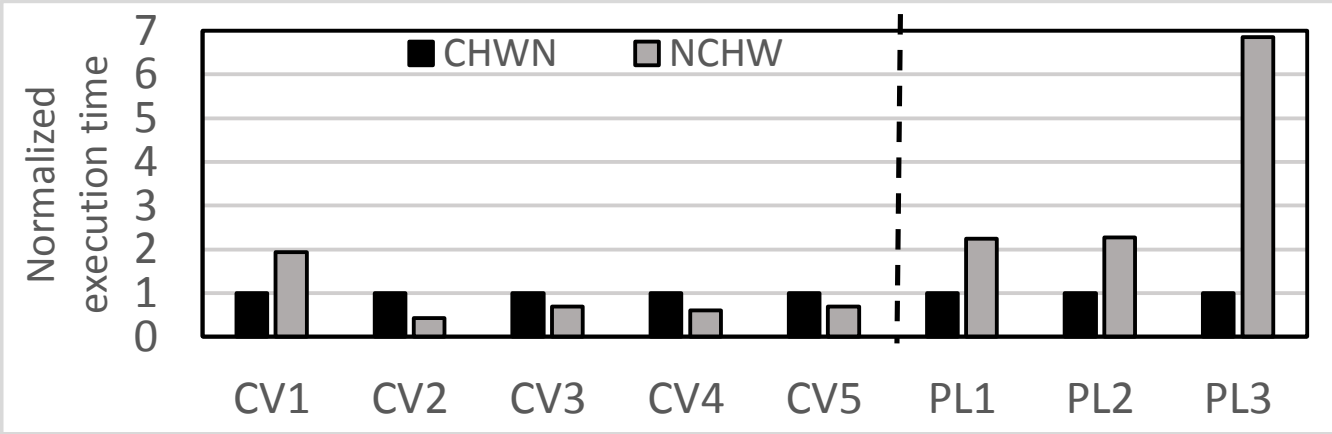

Fig. 1. Performance comparison between the CHWN layout (cuda-convnet2) and NCHW layout (cuDNNv4) on convolutional and pooling layers in AlexNet [12]

The first one is ***data layouts***: As the GPU thread organization, i.e., thread grid and thread block dimensions, is highly dependent upon data layout. Data layout determines the memory access pattern and has critical performance impact. For CNNs, the data are organized using multi-dimensional (4 dimensions) arrays. Depending on how we place data into different dimensions, we have many (24) ways to store the data in memory. While overlooked on previous work, surprisingly we found that the data layout can significantly affect the performance and memory efficiency. Fig. 1 shows the performance comparison of the two most popular data layouts on the AlexNet for different convolutional and pooling layers. From the figure, we can observe that up to 6.9x layer-level performance improvement could be retained by choosing a proper data layout. Moreover, even for the layers that have been always considered to be compute-bound, i.e., convolutional layers, we found that choosing the suitable data layout could lead up to 2.3x performance improvement. On the other hand, since each dimension has distinct memory access patterns and the size of each dimension can also affect the performance, the performance impact from data layout is complex and difficult for developers to reason about. The problem is further complicated when considering different types of layers in a CNN as each type may also prefer different data layouts. As shown in Fig.1, the different data layouts in both convolutional layers and pooling layers yield non-trivial performance differences, and a single data layout cannot deliver the best performance for all the layers. However, existing libraries only employ one data layout for all the CNN layers. Such a single uniform data layout in the existing design mismatches the inherent heterogeneity in different layers used in a CNN.

The second one is ***redundant off-chip memory accesses***. Our performance analysis shows that the memory efficiency of the memory-bounded pooling layers and classifier (i.e., softmax) layers is far from optimal due to the overlook on their off-chip memory data accesses. First, a CNN usually requires multiple steps to complete and there exists sequential data dependence across the steps. The common practice is to use a kernel for each step. However, it incurs high cost for inter-kernel data communication as the data pass through the bandwidth-limited off-chip memory. Second, leveraging data locality for high memory performance is an important optimization. However, how to optimize locality for different data layouts has not been addressed in existing CNN libraries.

In this paper, we look into these memory issues and propose a set of methods to optimize memory efficiency for accelerating CNNs on GPUs. The main contributions of this paper are:

- First, we characterize data layouts in various CNN layers, and reveal the performance impact of different layouts. Then we derive a light-weight heuristic to guide the data layout selection with minimal profiling overhead;
- Second, we support one network with multiple data layouts by proposing a fast multi-dimension data layout transformation on GPUs. We integrate the support for automatic data layout selection and transformation into a popular deep learning framework, Caffe.
- Third, we study the memory behavior of the memory-bounded pooling and softmax layers and optimize their memory access efficiency on GPUs.
- Finally, we perform rigorous evaluation and result analysis on different types of layers and representative networks, and demonstrate high performance improvements for both single layers, and complete networks.

With the promising results on the state-of-the-art networks including LeNet [17] and AlexNet [12], our work improves the development of deep neural network libraries on GPUs, hence contributing to the advance in machine learning applications.

## II. BACKGROUND

In this section, we first introduce the structure of a CNN, and summarize the algorithm characteristics for the major types of layers in CNNs. Then we describe the GPU-accelerated CNN libraries used in our study.

### *A. Convolutional Neural Networks*

CNNs are a type of forward feeding Artificial Neural Networks (ANNs) inspired from animal visual cortex organization. An example of CNN is shown in Fig. 2. As shown in the figure, the intermediate results are different sets of feature maps. The working principle of CNN is to extract the local features from high-resolution feature maps and combine them into more abstract low-resolution feature maps. These are realized by two alternating types of layers: convolutional and pooling layers. The last few layers are fully-connected classifiers that combine all local features together to produce the abstracted classification results.

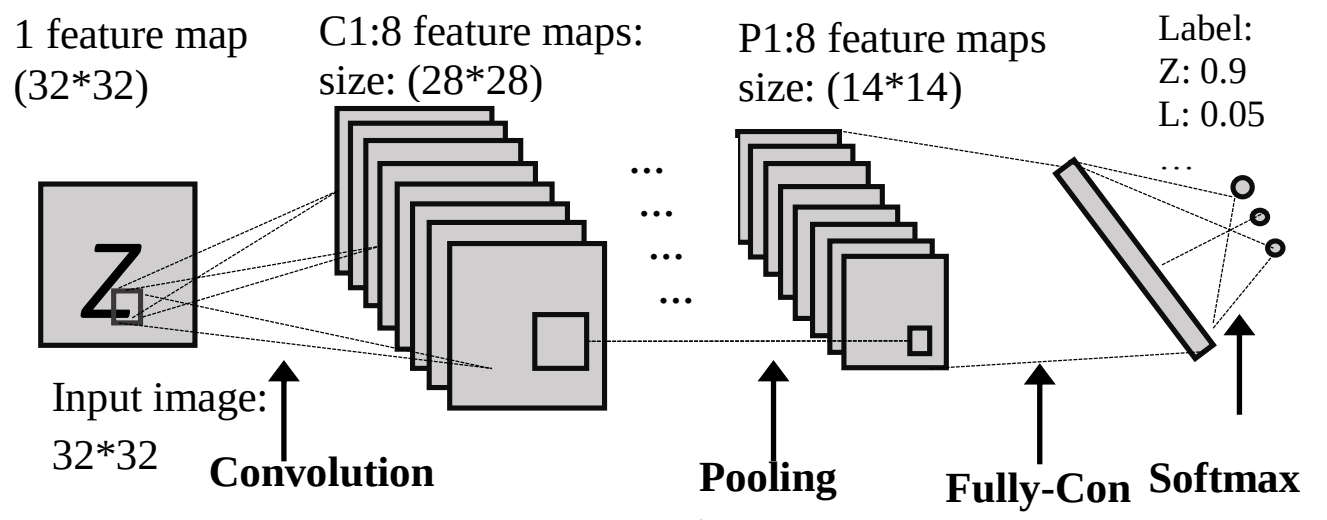

Fig. 2. The structure of an example CNN.

A ***Convolutional Layer*** extracts various features such as oriented edges, corners and crossings from input feature maps via convolutional filters, and then combines them into the more abstract output feature maps. The features in each feature map are 3D volume data with three dimensions: width, height and depth. With the large image data set and the massive computation power on GPUs, state-of-the-art CNN frameworks choose to process multiple images in a batch [12]. Thus, the input to a convolutional layer includes feature maps from multiple images and is organized as a four-dimensional (4D) array. The computation[1] in the convolution stage is shown in Equation 1, where ***Ni*** is the batch size, ***Ci*** is the depth or number of input feature maps, ***Hi*** and ***Wi*** are the height and width of a feature map, ***Fh*** and ***Fw*** represent the size of the convolution filter kernel, and ***Co*** is the output feature maps or the number of filters. The data volume in the convolution is four dimensional. To differentiate data layouts in the 4D arrays, we use the following notation in the paper: ***N*** (the number of images), ***C*** (the number of feature maps), ***H*** (the image height), and ***W*** (the image width). With this notation, we can see that Equation 1 uses the ***NCHW*** layout. In the ***NCHW*** data layout, the elements along the lowest dimension ***W*** are stored consecutively in memory. In comparison, the consecutive elements along the ***H*** dimension have a stride of ***W***; the consecutive elements along the ***C*** dimension have a stride of in ***H*W***; and so on.

$$Out_{co}[Ni][Co][Hi][Wi] = \sum_{Ci=0}^{C} \sum_{f_h=0}^{F_H} \sum_{f_w=0}^{F_W} In_{co}[Ni][Ci][Hi + f_h][Wi + f_w] * filter[Co][Ci][f_h][f_w] \quad \textbf{(1)}$$

Convolutional layers are typically most time consuming in a whole network. Therefore, achieving high arithmetic throughputs has been the main optimization objective [16][23]. Implementations using matrix multiplication and Fast Fourier Transform (FFT) have been proposed (see Section III.A). Counter-intuitively, we observe that the convolutional layers are not necessarily only compute bound. Specifically, for convolutional layers with small ***C*** and ***N*** dimensions, the performance is actually memory bound similar to 2D convolution (e.g., conv9 in Fig. 3). Furthermore, even for the compute-bounded layers, their memory organization especially the data layout will also have a substantial impact on their memory performance and overall performance. For instance, with a more suitable data layout, for the second convolutional layer in AlexNet, the utilization rate of ALUs (profiled on a

[1] The same data structure and convolution operation are used in both the forward pass and backward pass for testing and training CNNs [12].

NVIDIA Titan X GPU) can be improved significantly, from 55.64% to 78.71%.

A ***Pooling Layer,*** also called a down-sampling layer, summarizes the features from neighbors. In the example of Fig. 2, each feature is scaled down by a sampling factor of 2x2. The pooling layer can perform different operations such as average or max. An average pooling layer can be defined as Equation 2, where ***X*** and ***Y*** define the size of the pooling window, and ***stride*** is the distance between successive pooling windows. If the ***stride*** is smaller than the window size, the pooling is performed in an overlapped manner (i.e., overlapped pooling).

$$Out_{po}[Ni][Ci][Hi][Wi] = (\sum_{x=0}^{X}\sum_{y=0}^{Y} In_{po}[Ni][Ci][Hi*stride+y][Wi*stride+x])/Y/X \quad \textbf{(2)}$$

Pooling layers are usually paired with convolutional layers in CNNs. Compared to convolutional layers, pooling layers have low arithmetic complexity, *O(N*C*H*W)*. Its performance is mainly bounded by memory efficiency (i.e., bandwidth and latency).

A ***Classifier (Softmax) Layer*** is the final layer of a CNN for classification, which computes the possibility distribution over different labels. Before the softmax layer, there usually exist full-connected layers, which flatten the 4D feature maps into a 2D matrix. A standard matrix multiplication is used to implement a fully-connected layer [11]. The output will be fed into the softmax layer. The softmax layer will first find the maximal possibility over each batched image. Then, each possibility is shifted by the maximum. Next, an exponential operation is performed on each possibility. Last, all possibilities of each image are summed up and the summation is used to normalize the possibilities. The detailed algorithm is shown as the following five steps:

$$Maxv[Nx] = \sum_{x=0}^{X}\sum_{y=0}^{Y} \max(In[Nx][Cy]) \quad \text{//Step 1}$$

$$Midv1[Nx][Cy] = \sum_{x=0}^{X}\sum_{y=0}^{Y}(In[Nx][Cy] - Maxv[Nx]) \quad \text{//Step 2}$$

$$Midv2[Nx][Cy] = \sum_{x=0}^{X}\sum_{y=0}^{Y} \exp(Midv1[Nx][Cy]) \quad \text{//Step 3}$$

$$Sumv[Nx] = \sum_{x=0}^{X}\sum_{y=0}^{Y} sum(Midv2[Nx][Cy]) \quad \text{//Step 4}$$

$$Out[Nx][Cy] = \sum_{x=0}^{X}\sum_{y=0}^{Y}(Midv2[Nx][Cy]/Sumv[Nx]) \quad \text{//Step 5}$$

Each step in the softmax layer involves element-wise matrix or matrix-vector computation. The low arithmetic intensity in these matrix vector operations, and the intermediate data communication across different steps also make them memory bound.

### B. Deep CNN Libraries.

Till now, there are a number of frameworks [31] developed for CNN research. Among them, Caffe [11], cuda-convnet [12] and cuDNN [4], are most widely used and specifically optimized for GPU acceleration. In this paper, we study their memory efficiency for different types of layers. Caffe and cuDNN use the NCHW data layout for all the layers. There are two implementations for convolutions using this data layout. One is to use *Matrix Multiplication* (MM) to compute convolutions. This is the default approach as it can be used for different configurations of convolutional layers [4]. The other is based on *FFT*, which has been proposed and refined in recent works [19][23] and is integrated into the cuDNN version 4. The FFT method has significant overhead for intermediate data and works for certain types of convolution layers due to several limitations [23]. Different from Caffe and cuDNN, cuda-convnet chooses the CHWN data layout and uses the *Direct Convolution* method. For each library, we use their latest and best performing version, cuda-convnet2 and cuDNNv4.

## III. METHODOLOGY

In this section, we present our experimental methodology, including different types of benchmarking layers and networks, and the GPU hardware platform.

| Layer | Ni | Co | H/W | $F_w/F_h$ | Ci | S | Description |
|---|---|---|---|---|---|---|---|
| CONV1 (CV1) | 128 | 16 | 28 | 5 | 1 | 1 | LeNet[17]: Model Error rate: 0.18% (epoch 200) |
| CONV2 (CV2) | 128 | 16 | 14 | 5 | 16 | 1 | |
| POOL1 (PL1) | 128 | - | 28 | 2 | 16 | 2 | |
| POOL2 (PL2) | 128 | - | 14 | 2 | 16 | 2 | |
| CLASS1 | 128 images and 10 categories | | | | | | |
| CONV3 (CV3) | 128 | 64 | 24 | 5 | 3 | 1 | Cifar10[15]: Model Error rate:14.04% (epoch 100) |
| CONV4 (CV4) | 128 | 64 | 12 | 5 | 64 | 1 | |
| POOL3 (PL3) | 128 | - | 24 | 3 | 64 | 2 | |
| POOL4 (PL4) | 128 | - | 12 | 3 | 64 | 2 | |
| CLASS2 | 128 images and 10 categories | | | | | | |
| POOL5 (PL5) | 128 | - | 55 | 3 | 96 | 2 | ImageNet With AlexNet[12] Model |
| POOL6 (PL6) | 128 | - | 27 | 3 | 192 | 2 | |
| POOL7 (PL7) | 128 | - | 13 | 3 | 256 | 2 | |
| CLASS3 | 128 images and 1000 categories | | | | | | |
| CONV5 (CV5) | 64 | 96 | 224 | 3 | 3 | 2 | ImageNet with ZFNet Model[25] |
| CONV6 (CV6) | 64 | 256 | 55 | 5 | 96 | 2 | |
| CONV7 (CV7) | 64 | 384 | 13 | 3 | 256 | 1 | |
| CONV8 (CV8) | 64 | 384 | 13 | 3 | 384 | 1 | |
| POOL8 (PL8) | 64 | - | 110 | 3 | 96 | 2 | |
| POOL9 (PL9) | 64 | - | 26 | 3 | 256 | 2 | |
| POOL10 (PL10) | 64 | - | 13 | 3 | 256 | 2 | |
| CLASS4 | 64 images and 1000 categories | | | | | | |
| CONV9 (CV9) | 32 | 64 | 224 | 3 | 3 | 1 | ImageNet with VGG Model [22] |
| CONV10 (CV10) | 32 | 256 | 56 | 3 | 128 | 1 | |
| CONV11 (CV11) | 32 | 512 | 28 | 3 | 256 | 1 | |
| CONV12 (CV12) | 32 | 512 | 14 | 3 | 512 | 1 | |
| CLASS5 | 32 images and 1000 categories | | | | | | |

TABLE 1: THE CNNS AND THEIR LAYERS USED IN THE EXPERIMENTS.

### A. Benchmarks

The popular data sets for deep CNNs are MNIST [18], CIFAR [14], and ImageNet [21]. The MNIST data set is used for hand-written character recognition. It contains 50,000 handwritten digit images in the training set and 10,000 examples in the testing data set, and LeNet [17] is the best studied network for MNIST. CIFAR10 contains the 10 different categories of objects such as cat, truck, airplane, etc., and each category has 5000 training image and 1000 test images. We use the example Cifar network included in cuda-convnet for CIFAR10. ImageNet is a large-scale image collection with more than 1 million real-world images, which are classified into over 1000 categories. The commonly used benchmarking networks include AlexNet [12], ZFNet [25] and VGG [22]. AlexNet is widely used as the baseline network for exploring various new networks. These five networks cover a wide spectrum of problem sizes (from 28 to 256 image size) and network sizes. Table 1 presents the configurations for different types of benchmarking layers selected from the five networks. In this table, the convolutional layers have different configurations with various batch input sizes and feature map sizes; and the pooling layers include both overlapped and non-overlapped configurations. For softmax layers, we test twelve different configurations used in various classification problems (see Section VI).

*B. Experiment setup.*

We conduct experiments on a Linux machine with an Intel Xeon E5620 CPU. We use a NVIDIA GTX Titan Black GPU for all our tests. It contains 6144MB device memory, 5121 GFLOPS computing capability and 235GB/s effective memory bandwidth [30]. We also test our methods on a GTX Titan X which reports the similar trend, and the results are summarized in Section VI.

## IV. MEMORY ISSUE A: DATA LAYOUT

In this section, we first characterize the performance impact of data layouts in different types of CNN layers and derive the heuristics of selecting the suitable data layout based on the size and type of a layer. Then, we develop a fast multi-dimensional data layout transformation to support different data layouts in one network. Finally, we present how to apply our flexible data layout support into popular libraries/frameworks.

*A. Data Layout in Convolutional Layers*

Convolutional layers are the most critical layers for CNNs due to their dominant time in the network. As described in Section II.B, the implementation of a convolutional layer can be classified as generic ones including both the direct convolution and matrix multiplication method, and the special FFT method. We first look into the implication of data layout on generic ones. Then we analyze the effect of the FFT approach.

*Data Layouts in Generic Implementations*

For convolutional layers, we summarize their common properties from Table 1. First, the batch size, N, is generally a multiple of 16, and has limited choices as described in prior works [12][25]. Therefore, using N as the lowest dimension is a good choice to meet the requirements for coalesced memory accessing as the threads are organized accordingly. Given its limited choices, e.g., 32, 64 and 128, the optimization space is limited for this dimension. Second, although the width and height of each image typically are the same, the values can vary significantly for different convolutional layers, ranging from 12 to 224. Third, the depth of input feature maps (Ci) is 1 for grey-scale images or 3 for RGB images in the first convolutional layers, and then is a multiple of 16 in the rest of the convolutional layers, which can also provide regular memory accesses for a warp of GPU threads (warp size =32).

Based on the property discussed above on each dimension, combining the *W* and *H* dimension and using the dimension N as the lowest one are reasonable choices. This way, we have two candidate data layouts: CHWN or HWCN. The CHWN layout is used in cuda-convnet and each convolutional filter is applied on the *H* and *W* dimensions to generate one output feature map. A test of the HWCN layout shows the same performance as the CHWN layout on cuda-convnet because it doesn't change the memory coalescing feature for the *N* dimension and retains the same data reuse on the rest dimensions.

On the other hand, since the 2D convolution operations are applied on the H and W dimensions, a popular convolution implementation is to map the 4D matrix into a 2D array and perform computation in the form of matrix multiplication. The NCHW[1] data layout is used in this case and it is the strategy used in Caffe and cuDNN since matrix multiplication is a well-tuned algebra primitive available on virtually any platform with any input size. The special FFT implementation in cuDNN also inherits the NCHW layout and we analyze its distinct effect later in this section. Here we use cuDNN to denote its default MM method. As Caffe also binds cuDNN in its implementation as an improved version, our main comparison is between the CHWN (cuda-convnet) and NCHW (cuDNN) data layouts. In our following analysis, in the figures where cuDNN is always better than Caffe, we omit the results of Caffe.

Fig. 3 shows the performance comparison between two different data layouts in the convolutional layers. As discussed above, different data layouts will also impact on the implementations of the convolutional layers. The performance of each data layout is evaluated using their best performant implementations [4][15]. We can make the following observation. First, as shown in Fig. 3, cuda-convnet outperforms cuDNN for the CONV1 ~ CONV5 and CONV9 layers (by up to 6.5x speedup), but performs worse in the rest six convolutional layers. The differences between these two sets of layers lie in the ***N*** and ***C*** dimensions. Among the six layers performed better with cuda-convnet, CONV1, CONV3, CONV5 and CONV9 are the first layer in their corresponding CNNs, and the value of C of these layers is either 3 or 1. The layers, CONV2 and CONV4, also have a relatively small value of ***C***, no larger than 64. For the rest layers performing better with cuDNN, the values of ***N*** are either 64 or 32. To further identify the sensitivities of data layouts on each dimension, we collect the results with one varying dimension size (***N*** or ***C***) and the other three being fixed.

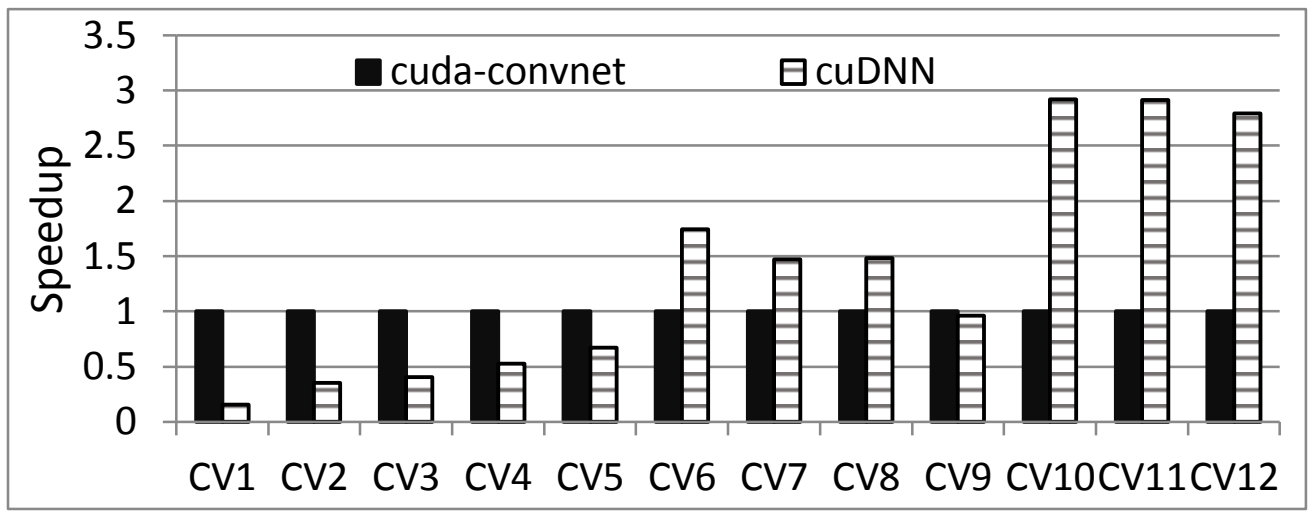

Fig. 3. Performance comparison between two different data layouts for the convolutional layers in Table 1. The performance is normalized to cuda-convnet measured on a GTX TITAN BLACK.

Fig. 4a shows the performance sensitivity when varying the value of ***N***. We can see that cuda-convnet with the CHWN data layout is more sensitive than cuDNN as the value of ***N*** changes. As shown in the figure, cuda-convnet outperforms cuDNN when the batch size *N* is more than 64. With the underlying CHWN data layout, Cuda-convnet first allocates a warp of 32 threads in a TB to process 32 images such that the memory accesses are coalesced. In order to further reduce off-chip memory accesses, if the batch size *N* is 128, cuda-convnet enables each thread to handle four images so that the data of these four images can be reused in the register file. If the number of images is less than 128, the reuse for images per thread would be reduced. As a result, the performance degrades quickly as the number of

1: cuDNN also supports the NHWC data layout and our tests show that its NCHW layout outperforms its NHWC layout.

images is reduced. In other words, for the CHWN data layout, the ***N*** dimension is used for both *memory coalescing* and *data reuse* (in registers), and therefore the performance is very sensitive to the value of ***N***.

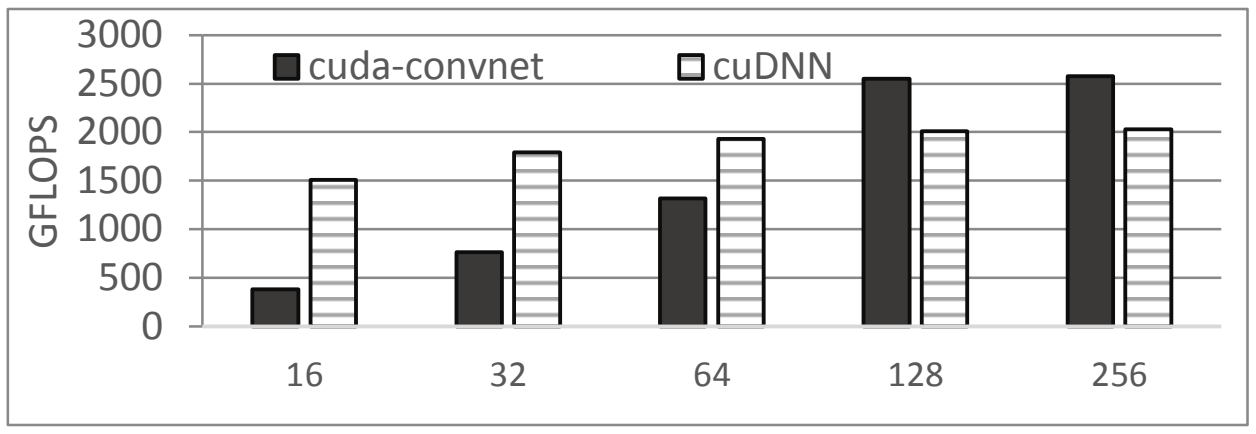

a. The performance with different values of N.

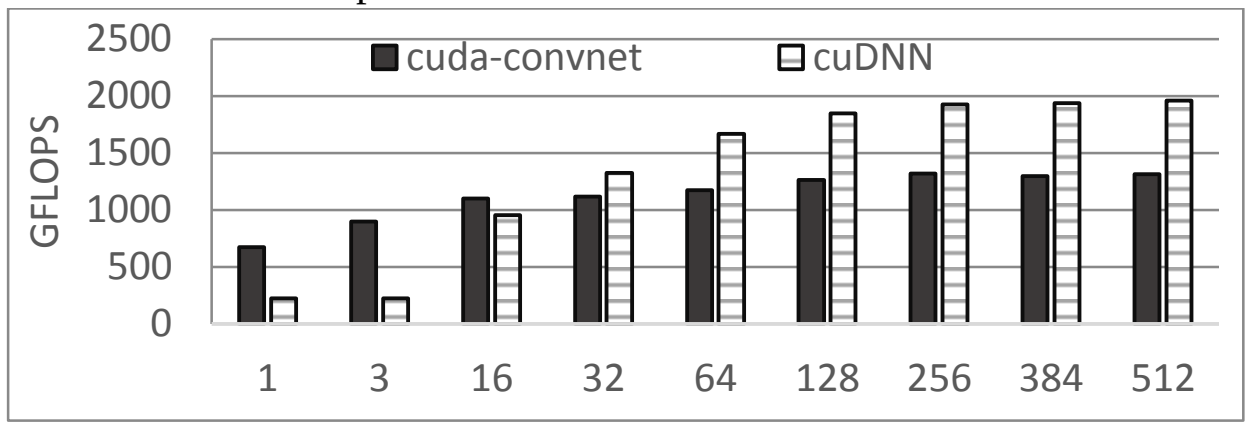

b. The performance with different values of C.

Fig. 4. Sensitivity study of data layouts on the N and C dimensions. CONV7 in Table 1 is used while others show similar trends.

Compared to cuda-convnet, cuDNN and Caffe use the NCHW layout, and utilizes the cuBLAS [27] library (or internal routine) for matrix operations. Since a matrix multiplication has only two dimensions, a matrix unroll step (along H and W) is needed to expand the input matrix, and merge multiple dimensions into two dimensions [11]. Such matrix transformation overhead is more evident when the matrix size is limited. As a result, as shown in Fig. 4b, when the value of ***C*** is less than 32, cuda-convnet (i.e., the CHWN layout) performs much better as it doesn't have the overhead of matrix expansion. On the other hand, cuDNN performs better when ***C*** is larger than 32 where matrix expansion leads to better data reuse and higher parallelism due to dimensions merging [4].

The performance implications from the data layout analysis above are two-folds. First, the ***N*** and ***C*** dimension are revealed as being highly correlated with memory performance. Second, a heuristic to select the suitable data layout for a convolutional shape can be derived based on the performance sensitivity analysis. For a given convolutional configuration, (1) if the value of *C* is smaller than a threshold *Ct*, CHWN will be preferred as the cost of memory transformation used in NCHW data layout is high; (2) if *N* is greater than or equal to a threshold *Nt*, the CHWN data layout is still the better choice as *N* is large enough to achieve both memory coalescing and data reuse. For the rest of the configurations, NCHW is the preferred choice. Due to different memory designs, including cache capacity and memory bandwidth, on different GPUs, the thresholds (*Ct* and *Nt*) can vary. For example, on the experimented Titan Black GPUs, the (*Ct*, *Nt*) is (32,128). On a new GPU such as GTX Titan X GPU, (*Ct*, *Nt*) is (128, 64). Considering that the heuristic parameters only relate to the property of the hardware, for each GPU architecture, we only need one-time profiling (as the one shown in Fig. 4 on varying N and C) to determine the thresholds.

*Data Layouts in FFT-based Implementations*

Besides the implementation of convolution using direct computation and matrix multiplication, convolution can also be computed in the frequency domain using FFT since convolution in the space/time domain is equivalent as element-wise multiplication in the frequency domain. The FFT-based method for implementing convolution has three steps: 1) transform the input feature image and filter kernel into the frequency domain using FFT; 2) perform element-wise product in the frequency domain; and 3) transform the result feature map from the frequency domain back into the space/time domain using inverse FFT. This approach has algorithmic advantage over matrix multiplication due to the lower computational complexity of FFT.

For FFT-based approaches for convolution, additional memory is needed to pad the filter kernel to match the size of feature maps and this overhead can be large for small filters. One way to reduce such overhead is to use tiling (i.e., FFT-tiling). The NCHW data layout has been used in the FFT-based approach [4][23].

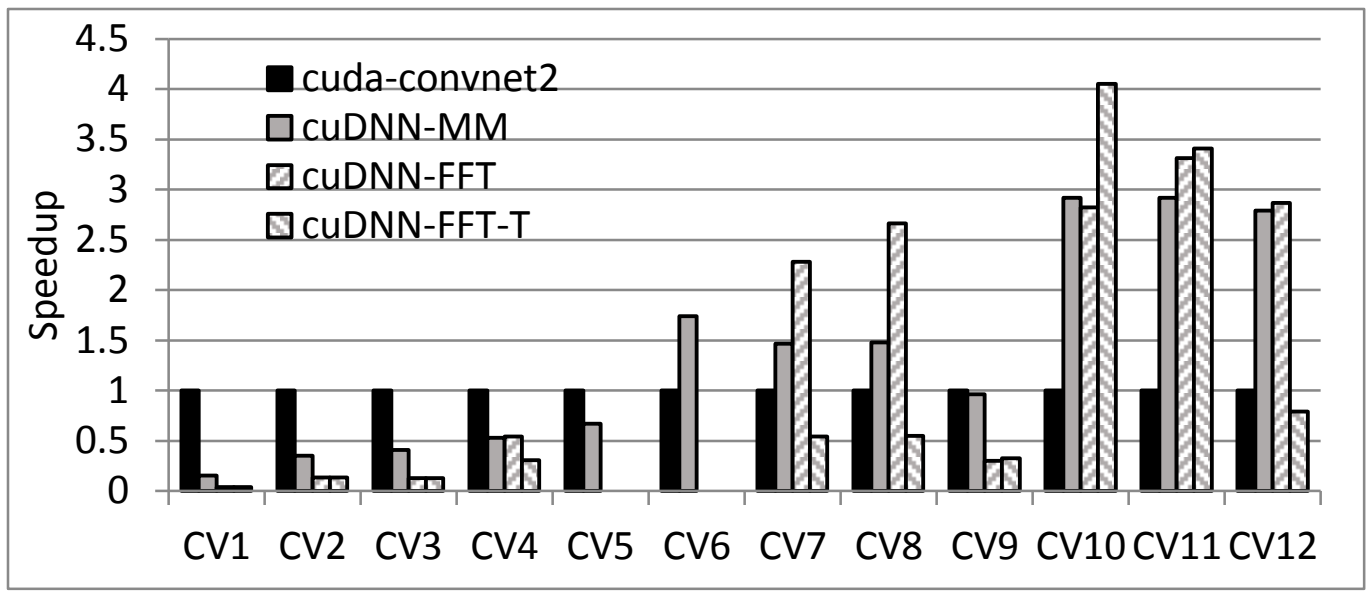

Fig. 5. Speedups of the FFT-based approach over the cuda-convets.

The latest cuDNN library version 4 provides two options for the FFT method: FFT and FFT-Tiling. The FFT option computes the convolution as described above but requires a significant memory for data padding and intermediate results. The FFT-Tiling option also uses FFT but splits the inputs into 32x32 tiles such that the memory overhead can be reduced compared to the FFT option. Fig. 5 shows the performance of various convolutional layers using FFT, FFT-Tiling and Matrix Multiplication (denoted as "cuDNN-FFT", "cuDNN-FFT-T" and "cuDNN-MM") with the NCHW layout compared to cuda-convnet with the CHWN data layout. For layers including CV5 and CV6, there are no results for both FFT options due to execution failures. For these layers, the required memory exceeds the hardware limit of 6GB on our GPU card. Among the layers that the FFT-based approaches can execute, there is no clear winner between cuDNN-MM and the cuDNN-FFT/cuDNN-Tiling method (they all use the NCHW layout). The FFT-based approach can perform better than cuDNN-MM when the filter kernel is large, the batch size is large or there are many channels such as CV7, CV10 layers. On the other hand, for small channel sizes, such as CV3, CV9, it performs much worse than the MM method. In these cases, the overhead of multiple steps, i.e., multiple kernel launches, streaming memory in and out multiple times, and zero-padding to the input size, in the FFT approach can outweigh its algorithmic advantage. Regarding the impact of data layouts, we can see that the FFT implementations do not change our observations made from Figure 5: the CHWN layout is preferred in CV1~CV5 and CV9 while the NCHW layout is preferred in other layers. Our data layout heuristic remains effective in selecting the suitable data layout for each convolutional layer.

### B. Data Layout In Pooling Layers

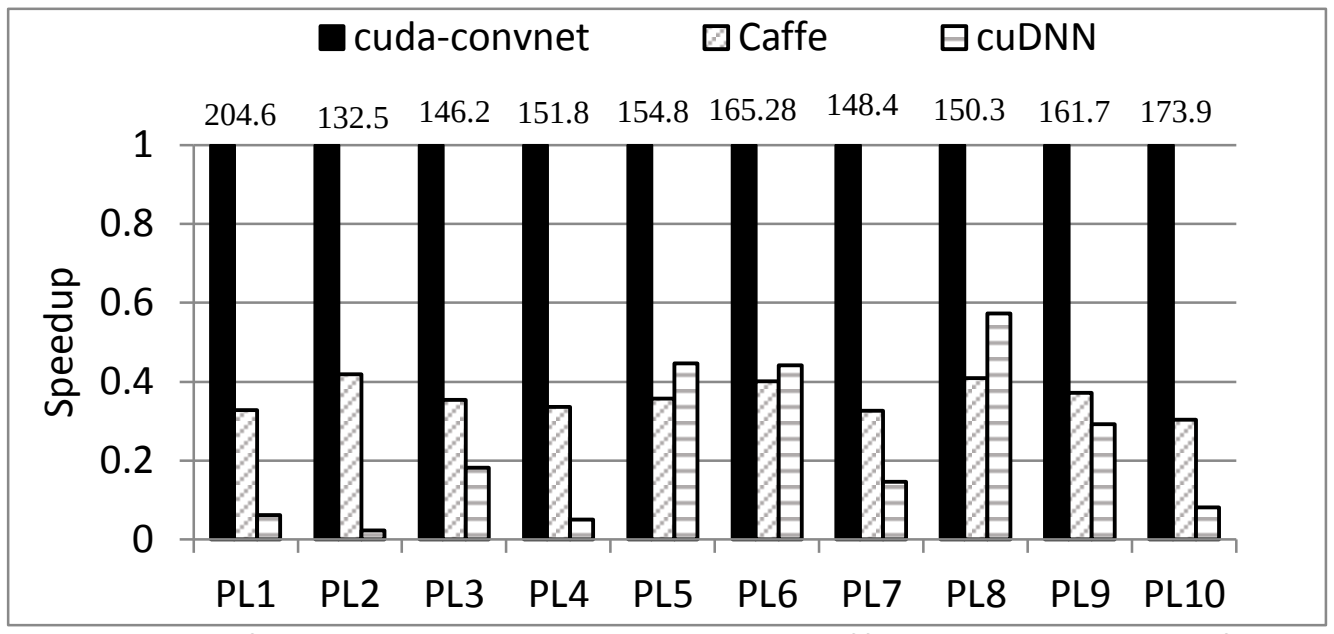

Figure 6. Performance comparison between different data layouts for the pooling layers in Table 1. The performance is normalized to cuda-convnet. The numbers on top denote the highest bandwidth (GB/S) achieved for each layer.

Compared to convolutional layers, pooling layers, another essential part of CNNs, are memory-intensive and also work on 4D data structures. Fig. 6 shows the performance of pooling layers with different data layouts. As we can see, cuda-convnet (i.e., CHWN) significantly outperforms Caffe and cuDNN (i.e., NCHW) across the board, with a speedup up to 16.3x. For the pooling operation on the CHWN data layout, it works through each slice of feature maps in the 4D array and memory coalesced accesses can be achieved along the lowest N dimension. However, for the NCHW data layout, the way of memory accesses is different. As the H and W dimensions are in the lowest dimensions, the pooling operations on each pooling regions of the feature map are directly applied to the pixels that are stored in memory consecutively. As shown in Equation 2, to compute an output element, each thread will access a pooling window of input elements. Therefore, the consecutive threads in a warp generate memory accesses with a stride. Such strided accesses from a warp are un-coalesced, resulting in over-fetching and poor memory efficiency. For this reason, for pooling layers, their memory access pattern determines that the CHWN layout is always preferred compared to the NCHW data layout, as shown in Fig. 6. When pooling is conducted in an overlapped way, the data layouts also impact on memory locality as there will be data reuse across overlapped pooling windows. Such locality impact is discussed in Section V.

### C. A Fast Data Layout Transformation for CNNs

From previous subsections, we reveal that a single data layout cannot satisfy the diverse layer configurations and layer types in a network. We also derive the preferred data layout based on the performance implication of the memory behavior of different layers. A subsequent question is how to enable the different suitable data layouts into one network? We propose an efficient data layout transformation. For brevity, we will mainly discuss the approach to transform from CHWN to NCHW.

Transforming an array in the CHWN layout to the NCHW layout is essentially a transpose operation on a 4D array. To implement parallel transpose for a 4D array on GPUs, a simple method is to construct a four dimensional thread hierarchy, and each thread dimension is used to handle a dimension of the array as shown in Fig. 7a. The issue of this implementation is that the memory accesses of writing into the output array is not coalesced as the threads in a warp have a long stride of CxHxW when accessing memory, causing severe bandwidth underutilization. To eliminate the un-coalesced memory accesses and achieve the optimized performance, we propose three optimizations as illustrated in Fig. 7b.

First, we observe that among two data layouts NCHW and CHWN, three dimensions including *C*, *H* and *W*, have the same relative positions. Thus, we combine these three dimensions into a single dimension as CHW. Then NCHW becomes [N][CxHxW], and CHWN becomes [CxHxW][N] after combination. This way, we downgrade the 4D transformation into 2D data layout transformation by flattening the matrix such that a 2D thread hierarchy can handle it, as shown in Line 4~5 in Figure 7b. Then we can apply the shared memory-based Tile optimization [30] to achieve the coalesced memory access for global writes in Line 7~14. Additionally, in the Kepler architecture, shared memory supports two access modes: 4-byte accesses and 8-byte accesses. To full utilize the bandwidth in 8-byte mode, we apply vectorization by grouping two consecutive float variables into a single vector type variable of *float2* to form a larger tile. When writing-back the tile, it can be scattered into multiple consecutive rows based on the tile shape, with each vector variable writing in a coalesced manner as shown in line 16-24. This extra vectorization can further boost the bandwidth utilization as the global access transactions will be doubled for data fetching. It is applied when N is larger than or equal to 64.

**Naïve Kernel (a)**

```
1.  __global__ void Transformation (float *in, float *out) {
2.  //from CHWN  to  NCHW.
3.   int tx =threadIdx.x, bx= blockIdx.x;
4.   int by =blockIdx.y, bz=blockIdx.z;
5.   out[(((tx*gridDim.z+bz)*gridDim.y+by)*gridDim.x)+bx]  =
6.         in[(((bz*gridDim.y+by)*gridDim.x)+bx)*blockdDim.x+tx];}
```

**Optimized Kernel (b)**

```
1.  template <int N, int C>
2.  __global__ void OptTransformation (float2 *in, float *out) {
3.  int tx =threadIdx.x, ty= threadIdx.y, bx =blockIdx.x,
    by=blockIdx.y;
4.  //1 Matrix flatten 4D to 2D: [C][H][W][N]->[C*H*W][N]
5.  int D2_W= N/2;   int D2_H = gridDim.y*gridDim.x*blockDim.x;
6.  //Shared Memory Tile for Subblock Transpose
7.      __shared__ float2   sh[C][33];      //Padding 1 float2.
8.  for (int i=0;i< N /64;i++) {  //handle 64 images every time
9.       int m = by*gridDim.x*blockDim.y+bx*blockDim.y+ty;
10. //2 Subgrouping in Shared Memory
11.      int D3_H = m/32; int D3_W  = m % 32;
12.      int index  = D3_W + D3_H*32;
13.      sh[ty][tx] = in[index*D2_W+tx+i*32];
14.      __syncthreads();
15. //3 Vector Transpose Index
16. if(C%32==0) {
17.     out[(2*ty+i*64)*D2_H+(bx)*32+tx] = sh[tx][ty].x;
18.     out[(2*ty+1+i*64)*D2_H+(bx)*32+tx]= sh[tx][ty].y; }
19. else if(C%16==0){
20.     out[(2*ty+i*64)*D2_H+bx*32+tx] = sh[tx][ty].x;
21.     out[(2*ty+1+i*64)*D2_H+bx*32+tx]= sh[tx][ty].y;
22.     out[(2*(ty+16)+i*64)*D2_H+bx*32+tx] = sh[tx][ty+16].x;
23.     out[(2*(ty+16)+1+i*64)*D2_H+bx*32+tx]=
    sh[tx][ty+16].y;}
24.      __syncthreads();
25.     }//end loop
26. }//end kernel
```

Fig. 7. Kernel code for data layout transformation

### D. *Wrap Up: Automatic CNN Data Layout Support*

Here, we present the integration of our data layout support, including both data layout selection and transformation, into existing CNN frameworks. By default, and as observed by library developers, for every data layout there is a preferred optimized implementation, e.g., for CHWN, Direct Convolution (DC) is better than Matrix Multiplication (MM) or Fast Fourier Transformation (FFT). Accordingly, to achieve better performance, when we use a data layout, we select the best implementation, i.e., DC for CHWN and MM or FFT for NCHW. In the deep learning frameworks such as Caffe or Cuda-convnet, each CNN has a configuration file that defines a network structure by specifying a stack of various layers. Each layer is specified with the layer type and the size of this layer for the input, output data tensor and weight kernel. Applying our data layout support requires two changes. The first is to add a new field in each convolutional and pooling layer to indicate the data layout choice. By scanning through the network once, the field in each layer is set for which data layout is desired based on our proposed heuristic. The second is at the runtime of training or testing the network, at the completion time of one layer, an additional check is inserted to determine whether a data layout transformation is needed before passing the output to the next layer. To determine the overhead of data layout transformation over the performance improvement obtained from the suitable data layout, *one-time* profiling can be applied to fine tune the data layout settings automatically. The profiling time overhead is relatively low (e.g., 395ms for AlexNet in a complete forward-backward profiling) when compared to millions of images using the same data layout. Finally, by comparing the data layout fields of the current layer and the next layer, if different, the transformation as discussed in Section IV.C will be performed. We decide not to fuse data layout transformation into the kernels for convolutional layers, because the different thread block configurations and memory access patterns lie between the convolution and 4D transpose kernels. The performance is adversely affected when they are combined.

Automatic layout transformation and selection relieves the burden upon the machine learning developers from analyzing various layer configurations and reasoning about their intricate GPU performance implications. The results in Section VI shows the effectiveness of our proposed support on various types of networks.

## V. Memory Issue B: Off-Chip Memory Accesses

In compliance with our efforts towards efficient memory access for CNNs, we look into their inherent memory behavior. Our analysis shows that the memory performance of memory-bounded pooling and softmax layers is far from optimal. In this section, we look into their memory behaviors and propose effective optimizations to improve their memory efficiency.

### A. *Memory Analysis and Optimization on Pooling Layers*

In Fig. 6, we report the highest bandwidth achieved in each pooling layer in the three libraries. As we can see, the bandwidth utilization is not high especially for the overlapped layers (i.e. when Fw>S) with a maximum of 173.9 GB/S and an average of 156.5 GB/S. For Caffe and cuDNN, the average bandwidth is 52.3GB/S and 41.9GB/S, respectively. The reason for the poorly achieved bandwidth by Caffe or cuDNN compared to cuda-convnet is the data layout. As discussed in Section IV.B, the NCHW layout in the pooling layers leads to strided memory accesses, resulting in low access efficiency. Another common reason for the relatively low bandwidth utilization among all three libraries is the significant redundant data accesses. We illustrate it using a pooling operation on 12 consecutive elements in one dimension. As shown in Fig. 8, the pooling window size is 4 and will slide with a stride of 2. Based on the pooling algorithm, each output element needs to load 4 input elements and totally 20 global memory accesses are required for the five outputs. Among these 20 global memory accesses, a number of them are redundant, as highlighted in the shaded ones in Fig.8. When the input is a 2D image, such redundant memory accesses will further increase.

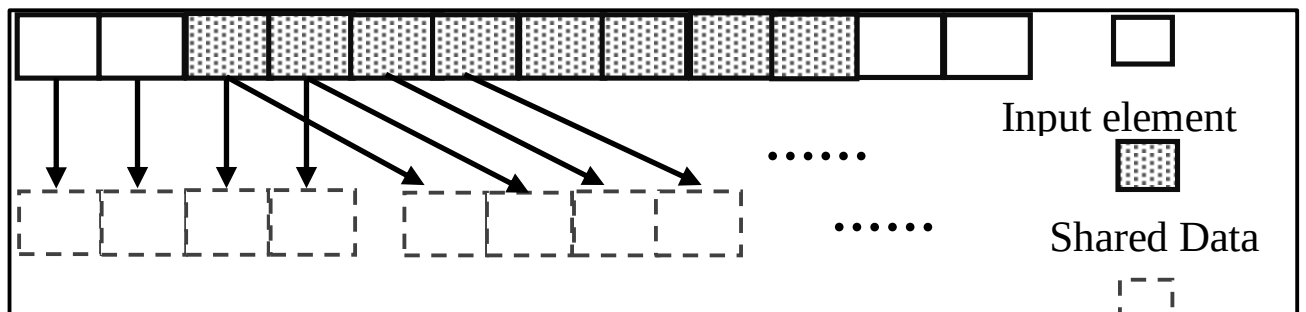

Fig. 8. Overlapped pooling with a window size of 4. The 2D image is simplified to 1D vector.

To achieve high memory efficiency for the pooling layers, the first optimization is to use the CHWN data layout. Then, we leverage on-chip register file to enable data reuse so as to reduce the off-chip memory requests. In the pooling layer, the on-chip working set can be defined as the number of the output elements. Therefore, when we expand the working set per thread (aka thread fusion/merge/coarsening [24]), we increase the number of the output elements per thread. Then within a thread, the input elements used to compute these output elements can be cached in the register file and only need to be loaded from off-chip memory once. To find the best working set expansion factors along both directions, we design an auto-tuning process which aims to balance the register pressure and data reuse with a fine-grain search. In order to converge into the optimal version quickly, we apply a hill-climbing heuristic to prune the search space. With an initial factor of 2, the expansion factor continues to increase linearly if the performance improves. Otherwise it stops as further expansion leads to high register pressure thus limiting the TLP and resulting in lower performance.

### B. *Memory Analysis and Optimization on Softmax Layers*

In Fig. 13 (see Section VI), we measured the highest bandwidth achieved for the softmax layers (the "BL_Best" bar) in existing libraries. As we can see, the overall memory performance is fairly low with the achieved highest bandwidth of 58.30 GB/s, less than 1/5 of the peak bandwidth. There are two main reasons. First, for the softmax layers, to ensure the data dependence across steps, cuda-convnet and Caffe use a separate kernel for each step and five kernels are used in total to implement the layer. Between consecutive kernels, the intermediate results are streamed in and out of global memory, resulting costly off-chip memory accesses. Second, in each step, there are two loops: one covers the batch size *N* and the other covers all categories. Since the outer loop, which covers all batched images, has no loop-carried dependency, Caffe and cuda-convnet parallelize it by allocating one thread for each iteration. However, the inner loop with loop-carried dependence is not parallelized, which is used to perform the reduction type operation (Section II.A) to compute the maximum or sum. The problem is that the parallelism of the outer loop is not enough for GPUs to hide instruction latency. If the number of images *N* is 128, very common for practical CNNs, the number of threads for the kernel is only 128.

To reduce the inter-kernel data communication cost, we propose to fuse all kernels into one such that the cross-step data communication can be promoted through registers or shared memory. For an efficient fusion, we need to consider three aspects: 1) different kernels may have different parallelism which needs to be coordinated for a uniform mapping into one kernel; 2) the synchronization and data dependence across steps need to be supported within the fused kernel; and 3) cross-step data communication needs to be enabled through fast on-chip memory accesses.

```
1.   dim3 threads(num_category);  dim3 blocks(num_img);
2.   __global__ void opt_kernel (float *mat, float *out){
3.   __shared__  float in_tile[C]; // C < 11K (k=1024)
4.   __shared__  float tmp_tile[1024]; //for  reduction
5.   int tidx = threadIdx.y*blockDim.xx+threadIdx.x;
6.   for(uint i = tidx;i<num_category;i=i+blockDim.y*blockDim.x)
7.        in_tile[i] = mat[blockIdx.x* num_category +threadIdx.x];
8.   // step 1
9.   max_reduction_thread_block (in_tile, tmp_tile);
10.  // step 2
11.  for(uint i = tidx;i<num_category;i=i+blockDim.y*blockDim.x)
12.       in_tile[i] = in_tile[i]-tmp_tile[0]; //tmp_tile[0] store the max
13.  ……}
```

Fig. 9. Optimized kernel after kernel fusion (C<11K)

To minimize the code change, we first identify that all five steps in the Softmax layer have the same two-level loops, and the implementation using five kernels also have the same TB configuration after parallelizing the outer loop. Therefore, we can fuse these five kernels into a single kernel without modifying the TB configuration. Second, since the output of a step is used as the input of its next step, the communication between two kernels becomes the inner-thread communication and the data used for the communication can be promoted into register file or shared memory (line 3-12 in Fig. 9). Thus, after kernel fusion, the intermediate global memory accesses are eliminated. Third, to address the problem of insufficient thread/memory-level parallelism, we propose to inject threads to further parallelize the inner loops. The inner loops of step 1 and 4 (Section II.A) perform reduction operations, while the inner loops of the rest steps have no loop-carried dependences across loop iterations. Since the reduction can be parallelized using shared memory and synchronization within a TB, we can enhance the parallelism in all five inner loops. The optimized kernel for the Softmax layers is shown in Fig. 9.

As shown in Fig. 9, the maximal value computed from step 1 can be stored in the shared memory and used in step 2 without using off-chip memory. Also, the input matrix elements can be loaded into shared memory or register in the first step and reused in the subsequent steps. The shared memory array, *tmp_tile*, is used as the intermediate temporary array during parallel reduction and *tmp_tile*[0] keeps the computed maximal value. In step 2 (line 10~11), we can reuse the elements in both shared memory arrays (i.e., *in_tile* and *tmp_tile*) to perform the matrix subtraction operations. Moreover, the shared maximal value can be further reused in the per-thread registers multiple times based on the multiple output elements each thread is to compute. As a result, inter-step memory communication is achieved now through the fast inter-thread shared memory coordination and intra-thread register reuse.

## VI. RESULTS AND ANALYSIS

In this section, we first evaluate each optimization for the single layers described in Table 1, and then evaluate the impact on all five complete networks. We experiment our optimizations in Caffe framework, where we inserted data layout selection and transformation code before the convolutional layer, and also adopted the best performing implementation of each convolutional layer for that data layout, e.g., we added the code using Direct Convolution and CHWN from Cuda-convnet, and invoked it if that data layout is selected. We also integrated our new implementations for Softmax and pooling layers as discussed in Section V into Caffe.

### *A. Results on Data Layout Optimization*

First, we show that data layout has significant performance impact on ***convolutional layers*** and our heuristics presented in Section IV can find the suitable layout for all convolutional layers in Table 1. By measuring the best performance that can be achieved on each data layout, Fig. 10 reports the performance differences, i.e., speedups, of the preferred data layout over the alternative (the bar labeled ‘Opt’). For example, on CV1, CHWN has an up to 6.5x speedup over NCHW; while on CV11, NCHW is the more suitable data layout, outperforming CHWN by 3.5x. On average, 2.48x speedup is achieved with the preferred data layout compared to use the alternative one. For the layers including CONV1, CONV2, CONV3, and CONV4, CHWN is the best layout as the value of N is 128. For the layers including CONV5 and CONV9, the number of input feature channels is less than 16. Thus, CHWN is still the best layout. For the rest layers, since the value of N is less than 128 and the value of C is more than 32, the NCHW layout achieves higher performance. Therefore, all the benchmarking layers in Table 1 confirm the effectiveness of our heuristics. We further examine the impact of our data layout optimization for various complete networks in Section VI.C.

Second, since consecutive layers in a network may have different preferred data layouts, the data layout transformation is needed and its overhead needs to be considered. Thus, we also evaluate the impact of the preferred data layout with the additional data layout transformation overhead. In Fig. 10, the performance bar labeled ‘Opt+Naïve Transform’ shows the speedup of the preferred layout with the overhead of a naïve transformation. The results labeled ‘Opt+Optimized Transform’ show the one with our optimized transformation. Their comparison highlights the impact of an efficient data layout transformation. For example, using the optimal data layout (i.e., CHWN) can provide 6.46x speedup for the CONV1. However, the overhead of the naïve transformation to achieve this data layout is large enough such that the data layout benefit is eliminated. With our optimized transformation, however, this layer still achieves 4.02x speedup. The exceptions are CONV9 and CONV5, whose convolution filter is small and performance difference is minor (only 4.1%) among different data layouts. Therefore, even using our optimized transformation, we cannot improve their performance. For these very small convolutions, our data layout transformation has very little negative impact as the layer itself has very minor impact on the overall networks (less than 5%). Overall, by considering the data layout transformation overhead on different layers, a naïve data layout transformation cannot sustain the significant performance benefit from the suitable data layout and may even degrade the overall performance. Using our proposed fast transformation to enable the optimal layout, an up to 4.02x (an average of 2.08x) speedup is achieved.

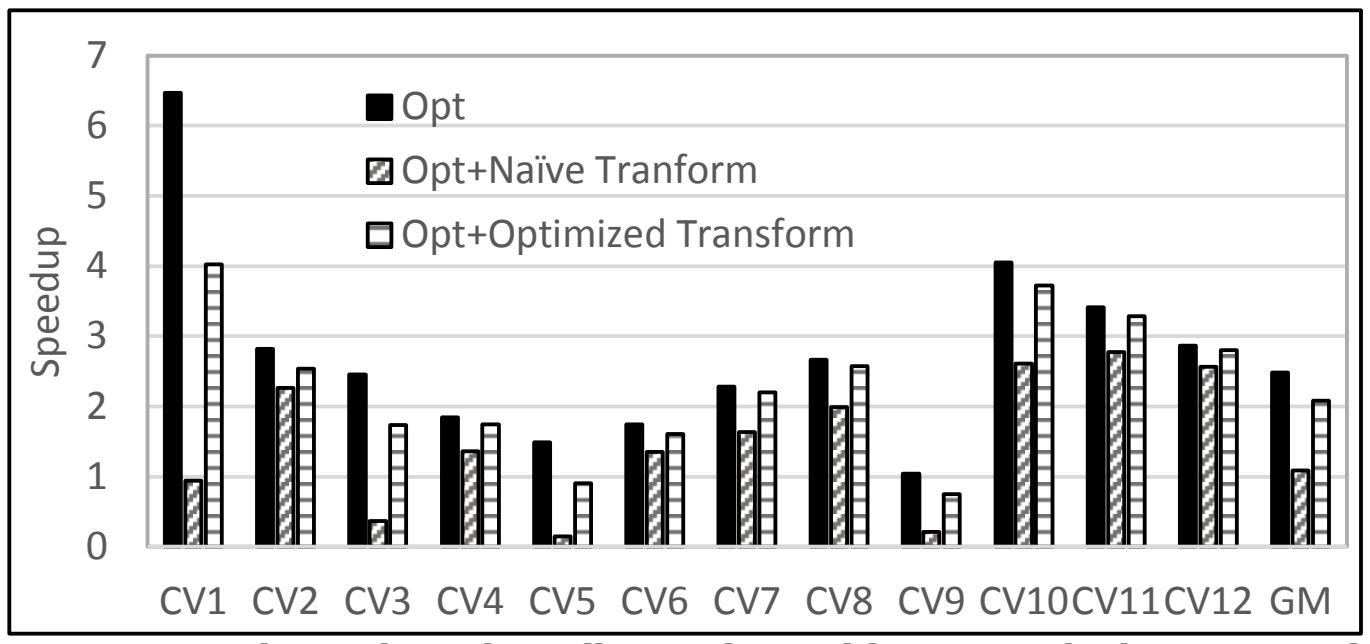

Fig. 10. Speedups achieved on all convolutional layers. For both NCHW and CHWN data layouts, the best achieved performance is measured to calculate the performance differences.

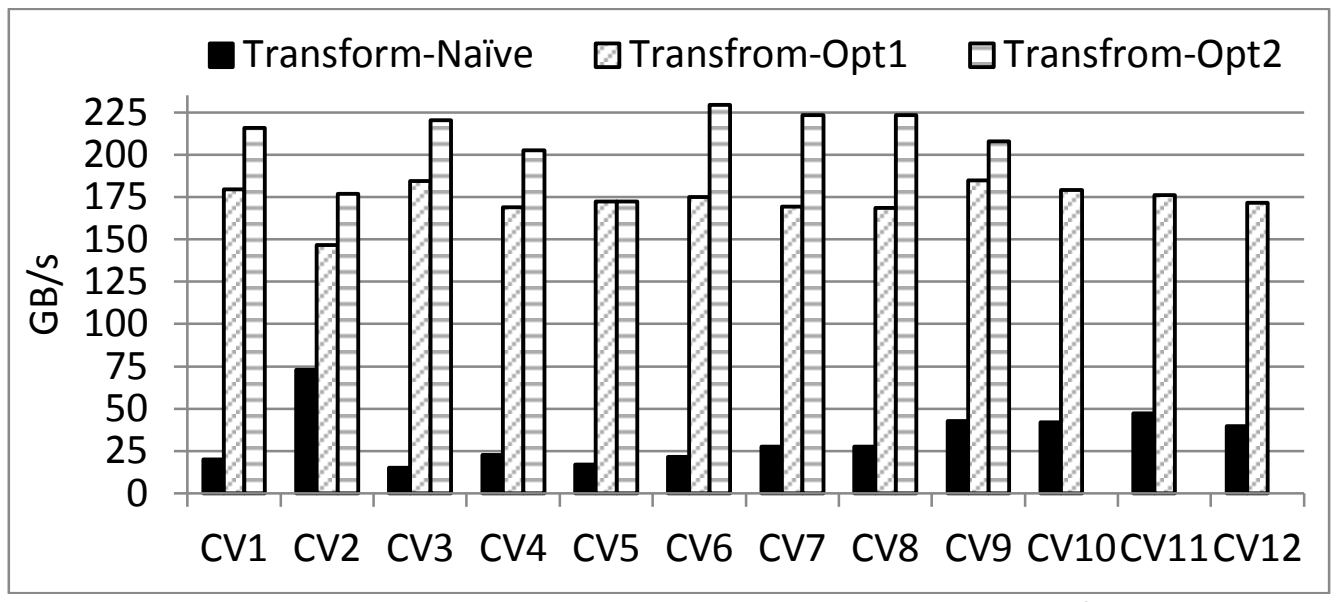

Fig. 11. Achieved memory bandwidth using three methods for data layout transformation. The Transform-Opt2 is not applicable for CV10, CV11, CV12 whose N is smaller than 64.

Fig. 11 shows the detailed performance evaluation of our proposed data layout transformations. The bar of "Transform-Opt1" applies layout flattening and tiling with shared memory transpose (Section IV.C). It significantly improves the performance with an average of 6.48x speedup for all type of layers. By further applying the vectoring technique (labeled 'Transform-Opt2') on the applicable layers (i.e., for those with N is over or equal to 64), the achieved bandwidth has been improved to up to 14.7x, and an average speedup of 7.5x. The optimized bandwidth for CONV6 has achieved of 229.5GB/S, which is 97.6% of the effective GPU memory bandwidth. Besides, the memory overhead in the transformation is also very low compared to the overall memory footprint. For instance, in AlexNet, the additional memory space overhead is only 73.5MB, which is less than 3% compared to the memory footprint of around 3GB. Furthermore, the additional memory, i.e., the input matrix (less than 3% overhead), is freed right after the layout transformation is completed.

## *B. Results on Off-chip Memory Access Optimization*

Fig. 12 shows the performance comparison of different ***pooling layers*** between the existing implementations and our optimized kernels, 'Opt', generated through auto-tuning. First, cuda-convnet outperforms the Caffe and cuDNN across the board, highlighting that the preferred data layout in pooling layers is CHWN. Second, with the preferred data layout of CHWN, for the overlapped pooling layer, our optimization on data locality labeled as 'Opt' can achieve higher performance with an average of 193.8GB/S memory bandwidth and improve the state-of-the-art performance by an average of 14.3%. This is the direct result of the significantly reduced global memory accesses through better data reuse. For example, in PL3 with a pool window of 3 and a stride of 2, our optimized kernel effectively reduced 9.1% global memory transactions and 36.0% DRAM accesses respectively, compared to cuda-convnet, and the overall performance has improved by 33.9%.

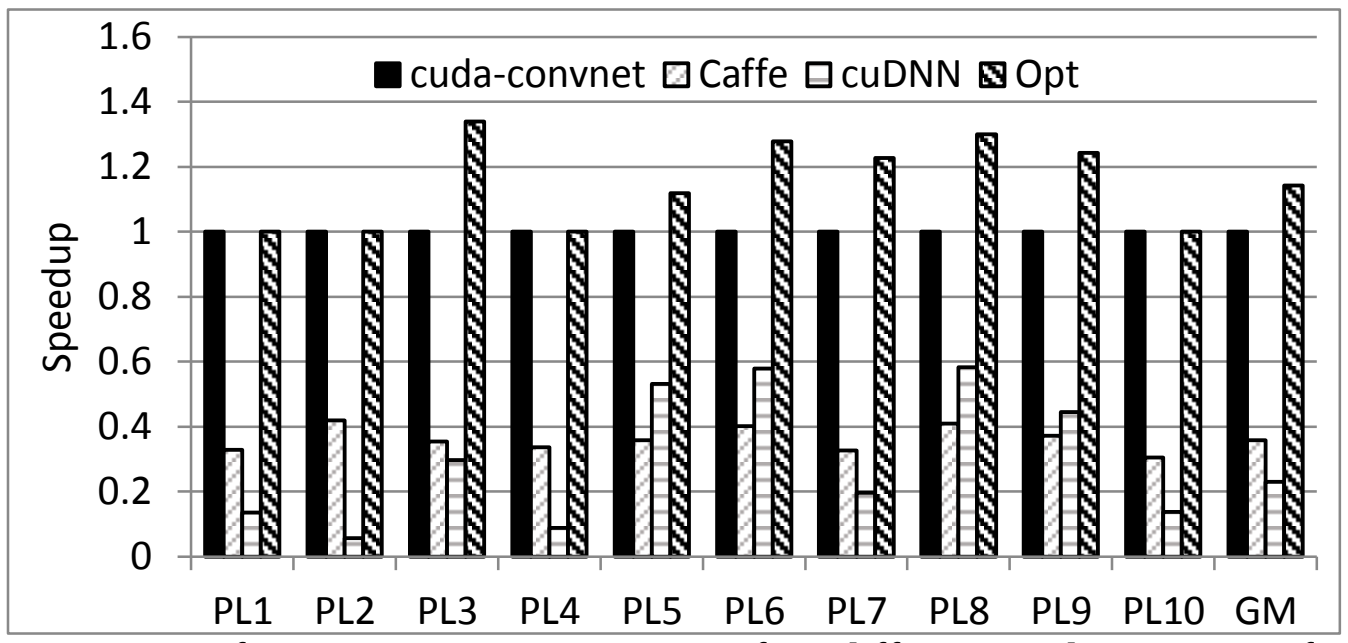

Fig. 12. Performance comparison among four different implementations for the pooling layers in Table 1. The performance is normalized to cuda-convnet.

Fig. 13 shows the memory bandwidth comparison between optimized and original kernels for the ***softmax layers***. The bar "BL_Best" shows the highest bandwidth achieved in existing libraries while the "Opt" shows the performance achieved using our optimized fused kernel. From the figure, we can see that our optimized versions consistently improve the memory bandwidth across all the softmax layers. For small layer sizes, the bandwidth cannot be well utilized. When the layer has large categories (such as 10000), the bandwidth achieved in "Opt" can reach 220.95GB/S, which is 94.02% of the effective GPU memory bandwidth. In comparison, the highest bandwidth achieved in the BL_Best implementation (cuDNN) is still fairly low, 58.30GB/S, due to the inefficient memory access behavior. Among the optimizations, the efficient inter-step data communication enabled by our kernel fusion has contributed up to 3.53x speedup and an average of 2.81x speedup using the geometric mean. More threads from parallelizing the inner-loops can further bring an average speedup of 5.13x.

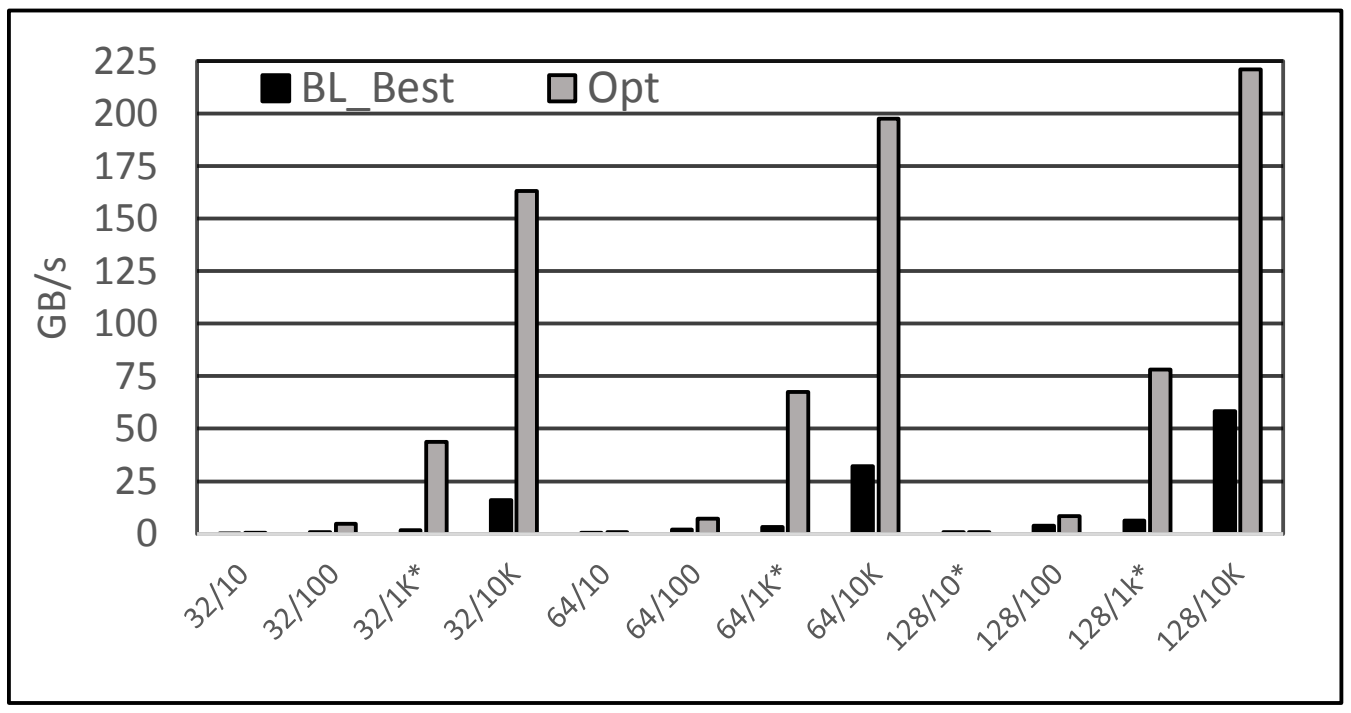

Fig. 13. Performance comparison (GB/S) of softmax layers with a wide range of configurations. x/y means the batch size as x and the number of categories as y.

### C. Results on Whole Networks

In this section, we evaluate the efficacy of our memory efficiency optimizations on the whole networks. We use Caffe as our base framework and bind our two memory optimizations: flexible data layout support and optimized memory accessing for pooling and softmax layers. The cuDNN library is also bound into Caffe, and runs as an improved version of Caffe. As cuDNN provides multiple implementation modes for the convolutional layer, each of them is performed to obtain the best performance. We name each of the evaluated mechanisms as follows:

- cuDNN-MM: the convolutional layers use the standard matrix multiplication mode in cuDNN.
- cuDNN-FFT: the convolutional layers use the FFT mode and falls back to the cuDNN-MM mode if failed.
- cuDNN-FFT-T: the convolutional layers use the FFT-Tiling mode and falls back to the cuDNN-MM mode if failed.
- cuDNN-Best: ***cherry-pick*** the best-performed/fastest one for each convolutional layer from all these available running modes.
- Cuda-convnet: the network is running on cuda-convnet.
- Opt: the network is running on the optimized framework using our proposed memory optimizations.

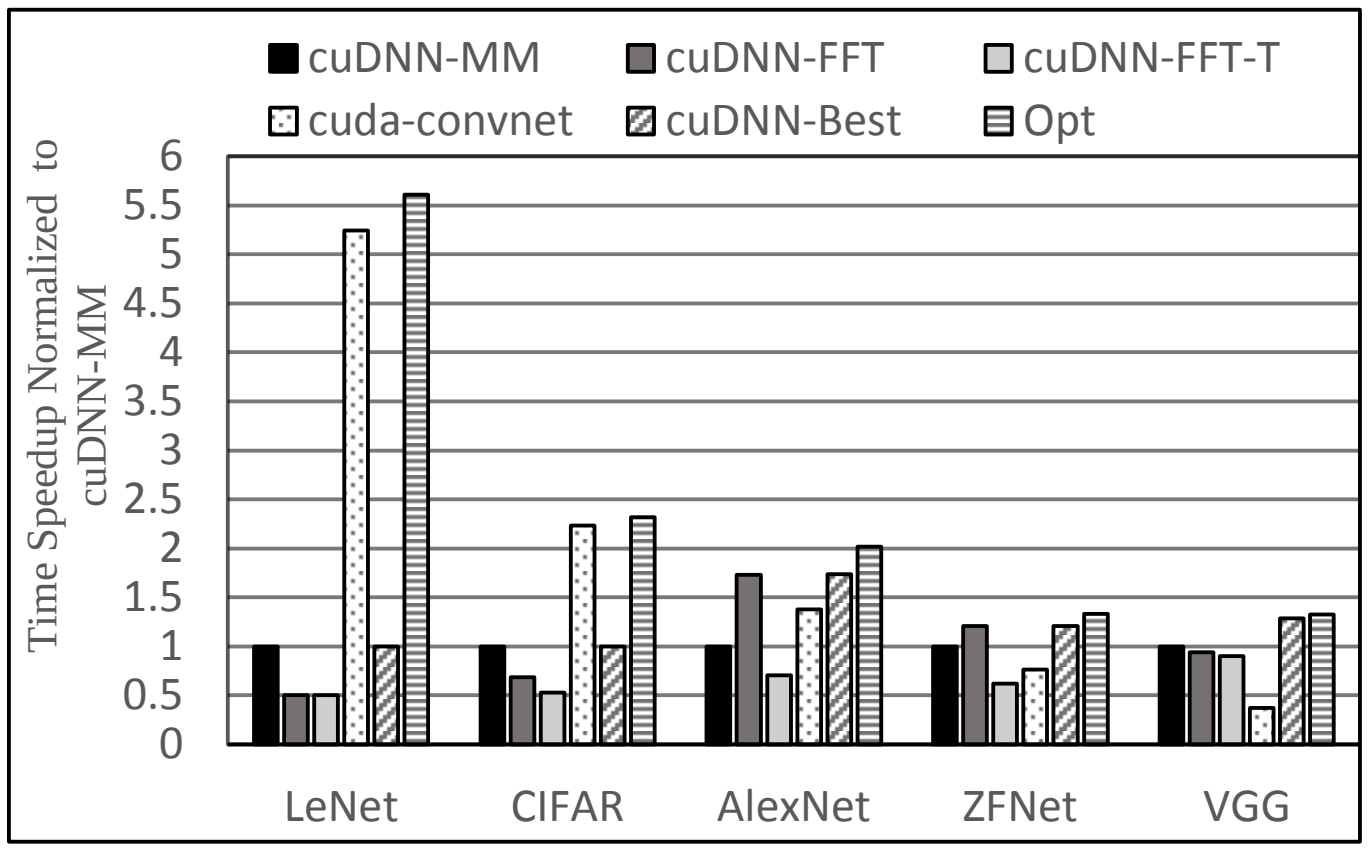

Fig. 14. The overall network performance comparison among various schemes.

Fig. 14 shows the overall execution time of the five complete CNNs using these different mechanisms. As we can see, with a single and uniform data layout in existing libraries, either cuDNN or cuda-convnet can only deliver the high performance for a subset of neural networks. For LeNet and Cifar, the performance of cuDNN is much worse than cuda-convnet, even using the best-possible performance from cuDNN-Best. On the other hand, cuda-convnet is significantly under-performed compared to cuDNN for AlexNet, ZFNet and VGG. This highlights the non-trivial performance drawbacks of employing fixed data layouts in current CNN libraries. By augmenting the flexible data layout and low-overhead layout transformation, our optimized framework can achieve the highest performance for all these networks. Compared to the state-of-the-art cuDNN library, considering the two most representative CNNs (LeNet and AlexNet), for LeNet, we can achieve 5.61x speedup over cuDNN-MM, 11.28x over cuDNN-FFT, 11.28x over cuDNN-FFT-T, and 5.61x over cuDNN-Best; for AlexNet, we can achieve 2.02x over cuDNN-MM, 1.17x over cuDNN-FFT, 2.87x over cuDNN-FFT-T and 1.16x speedup over cuDNN-Best.

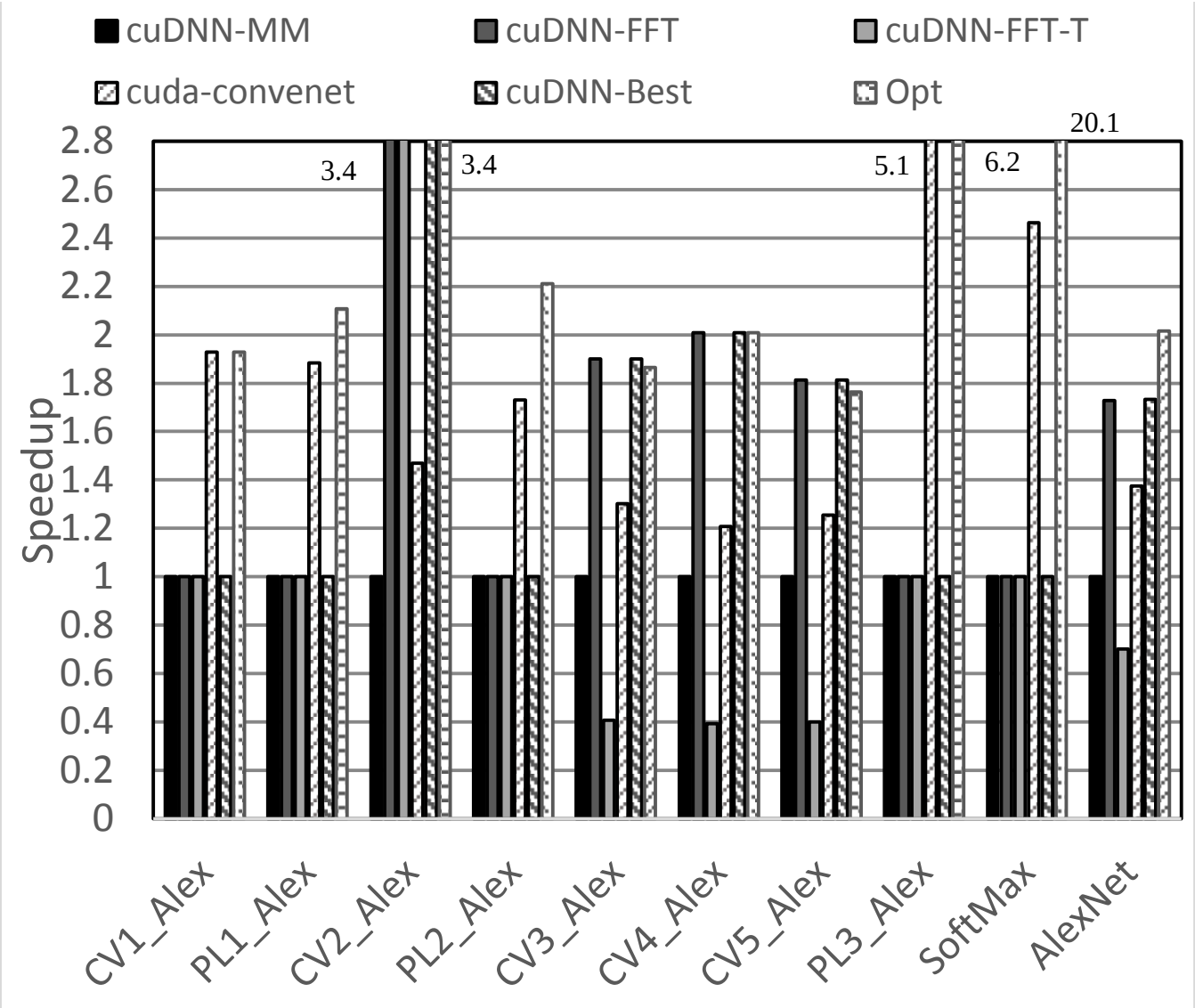

Fig. 15. The performance comparison of different layers in AlexNet, The performance is normalized to cuDNN-MM.

Fig.15 shows the detailed performance comparison of different layers in AlexNet. AlexNet is the de-facto deep CNN structure for machine learning research. It contains 5 convolutional layers, 3 pooling layers, 1 softmax layer, and other layers such as normalization and full-connected layers. The convolutional layers in AlexNet have different configurations and no single data layout can be suitable for the whole network. For CV1, cuda-convnet performs better than cuDNN due to the suitable data layout of CHWN, while for the remaining four convolutional layers, cuDNN outperforms cuda-convnet significantly. Our optimized framework selects the right layout across all different convolutional shapes, CHWN for CV1 and NCHW for the rest. Then, for the pooling layers, based on our study, CHWN is the best and cuda-convnet consistently performs better than cuDNN. Our optimizations on these three pooling layers further improves the performance by up to 27.8% over cuda-convenet. Finally, for the softmax layer, our memory optimization shows the significant speedup, with up to 20.1x speedup over cuDNN and 8.2x over cuda-convnet. As there are four data layout transformations happening after PL1, CV2, CV3 and CV5, only minor overhead is incurred as aresult of the efficient data layout transformation. Overall including the layout transformation overheads, our optimized frameworks improve AlexNet by 1.16x over cuda-convnet, and 1.46x over cuDNN-Best. The performance impact of each layer on the whole network is different, with convolutional layer being the most performance dominant. Thus, achieving the flexible data layout for a network is the most critical optimization, contributing a 72% improvement. Comparatively, the off-chip memory access optimization contributes 28% due to the much smaller execution time of pooling and Softmax layers.

Finally, our test on the NVIDIA Titan X (Maxwell) GPU shows the very similar trends. For example, compared to cuda-convnet, Caffe and cuDNN, our proposed optimizations achieve 1.04x, 24.5x and 11.84x speedup for the small network of MNIST; 5.11x, 1.77x and 1.05x speedup for a large network of VGG Net.

## VII. Related Works

With the state-of-the-art recognition accuracy, deep CNNs have been applied into numerous application domains including image recognition, speech translation, drug discovery, etc. The pervasive usage of CNNs has also ignited the research on how to improve the speed of CNNs with increasingly larger data sets because the training time is still the limiting factor for applying CNNs. Specialized accelerators [2][3] have been proposed to build high-performance memory logics for CNNs. However, these accelerators are hard to program as they only support limited function units for a subset of layers. GPU clusters [6] have been designed to achieve the high computational throughput. Their focus on performance optimization for multi-GPU CNNs [6][13] is how to efficiently partition the workload with low-overhead inter-device communication. Within each GPU node, the main execution blocks are still the basic layers as discussed in this paper. The improved single-node performance will certainly enhance the overall throughput of large scale systems.

To accelerate CNNs, GPU-based implementations for various types of layers such as convolutional, pooling, and softmax layers are available in various libraries [1][4][7][11][12] such as cuDNN and cuda-convnet. These GPU-based libraries relieve the significant burden in developing high performance CUDA code from machine learning researchers, and these GPU implementations have been widely used in the machine learning community. To accelerate the CNN performance on GPUs, recent efforts [16][23] focus on improving the computational efficiency of the convolutional layers while the data layout across the network has been unexplored. Cirensan et al. [5] implemented parallel convolution layers. Ren et al. [20] recently presented the vectorization of the basic layers. Marc [9] recently studied the coarse-grain parallelization using batch-level parallelism in CNNs. None focuses on the effect of the underlying data layouts and data access patterns for all these different CNN layers. To the best of our knowledge, this paper is the first study to look into the memory efficiency of accelerating CNN on GPUs. With the intricate data structure in CNNs, the performance implication of their memory behavior is complex. Our performance analysis unveils various performance implications and systematically optimize their memory efficiency, showing it to be a substantial performance factor for different types of CNNs.

We also observe that like the FFT approach, more techniques leveraging arithmetic complexity may be proposed in the future for CNNs, e.g., the recent proposal from Nervana Systems [16]. They can set state-of-the-art performance for a group of layers, for which they suit. On the other hand, the GPU hardware also continues to evolve quickly, such as the latest NVIDIA Pascal architecture, that begins to support FP16 (e.g., NVIDIA Tesla P100[33]) to enhance the computational throughput and reduce the memory usage significantly. Nevertheless, the underlying impact from data layout remains. The reason is that with compute efficiency being addressed with these new approaches, the performance impact of the memory efficiency is likely to become more important.

Since our optimizations and observations are based on memory efficiency, which is very important for this domain of deep learning, we believe that other accelerators, such as FPGAs [26] or ASIC-based processors [3], e.g., Tensor-Processing Unit (TPU) [32], may also benefit from more efficient memory access patterns. Therefore, we expect that our technique can be applicable for such systems.

## VIII. Conclusions

This work looks into the memory efficiency issues of current GPU-accelerated deep CNN implementations, including both data layouts and off-chip memory accesses. Our detailed study unveils the impact of data layouts on different types of CNN layers and their performance implications. Then, we propose efficient data layout support as our solution. We further look into the memory access patterns of the memory-bounded layers, and propose effective optimizations to substantially reduce their off-chip memory requests and inter-kernel communication. The experiments demonstrate the effectiveness of our memory optimizations and their universal effects on different types of layers and various complete networks.

## IX. Acknowledgement

We thank the anonymous reviewers for their valuable comments to improve our paper. This work is supported in part by NSF grants CCF-1216569, CCF-1618509, a Chinese research program "introducing talents of discipline to universities B13043", and an AMD gift fund. We also want to thank Amro Awad for his insightful suggestion and generous help on improving the work.

## References

[1] James Bergstra, Olivier Breuleux, Frederic Bastien, Pascal Lamblin, Razvan Pascanu, Guillaume Desjardins, Joseph Turian, David Warde-Farley and Yoshua Bengio. Theano: A CPU and GPU math compiler in Python. In SCIPY, 2010.

[2] Srimat Chakradhar, Murugan Sankaradas, Venkata Jakkula and Srihari Cadambi. A dynamic configurable coprocessor for convolutional neural networks. In ISCA, 2010.

[3] Tianshi Chen, Zidong Du, Ninghui Sun, Jia Wang, Chenyong Wu and Yunji Chen. DianNao: A Small-footprint high-throughput accelerator for ubiquitous machine-learning. In ASPLOS, 2014.

[4] Sharan Chetlur, Cliff Woolley, Philippe Vandermersch, Jonanthan Cohen, John Tran, Bryan Catanzaro and Evan Shelhamer. cuDNN: Efficient primitives for deep learning. CoRR, abs/1410.0759.

[5] Dan C. Ciresan, Ueli Meier, Jonathan Masci, Luca M. Gambardella and Jurgen Schmidhuber. Flexible, high performance convolutional neural networks for image classification. In IJCAI, 2011.

[6] Adam Coates, Brody Huval, Tao Wang, David J. Wu and Andrew Y. Ng. Deep learning with COTS HPC systems. In ICML, 2013.

[7] Ronan Collobert, Koray Kavakcuoglu and Clement Farabet. Torch7: A matlab-like environment for machine learning. In NIPSW, 2011.

[8] Jason Cong and Bingjun Bao. Minimizing computation in convolutional neural networks. In ICANN 2014.

[9] Marc Gonzalez Tallada. Coarse Grain Parallelization of Deep Neural Networks. In PPoPP, 2016.

[10] Kaiming He, Xiangyu Zhang, Shaoqing Ren and Jian Sun. Delving deep into rectifiers: surpassing human-level performance on ImageNet classification. CorRR, abs/1502.01852.

[11] Yangqing Jia, Evan Shelhamer, Jeff Donahue, Sergey Karayev, Jonathan Long, Ross Girshick, Sergio Guadarrama and Trevor Darrel. Caffe: convolutional architecture for fast feature embedding. CoRR, abs/1408.5093, 2014.

[12] Alex Krizhevsky, Ilya Sutskeve and Geoffrey E. Hinton. ImageNet classification with deep concovlutional Neural Networks. In NIPS, 2012.

[13] Alex Krizhevsky. One weird trick for parallelizing convolutional neural networks. CoRR, abs/1404.5997, 2014.

[14] Alex Krizhevsky. Learning multiple layers of features from tiny images. Technical Report, 2009.

[15] Alex Krizhevsky. cudaconvet2: http://code.google.com/p/cuda-convnet2/ , 2014.

[16] Andrew Lavin and Scott Gray. Fast Algorithms For Convolutional Neural Netoworks. Arxiv. CoRR, abs/1509.09308, 2015.

[17] Yann LeCun, Leou Bottou, Yoshua Bengio and Patrick Haffner. Gradient-based learning applied to document recognition. In Proceedings of the IEEE, 1998.

[18] Yann LeCun, Corinna Cortes, and Christopher J.C. Burges. The mnist database of handwritten digits. http://yann.lecun.com/exdb/mnist/, 2015.

[19] Michael Mathieu, Mikael Henaff and Yann LeCun. Fast training of convolutional networks through ffts. CoRR, abs/1312.5851.

[20] Jimmy SJ. Ren and Li Xu. On vectorization of deep convolutional neural netowrks for vision tasks. In AAAI, 2015.

[21] Olga Russakovsky, Jia Deng, Hao Su, Jonathan Krause, Sanjeev Satheesh, Sean Ma, Zhiheng Huang, Andrej Karpathy, Aditya Khosla, Michael Bernstein, Alexander C. Berg and Li Fei-Fei. ImageNet Large Scale Visual Recognition Challenge. In IJCV, 2015

[22] Karen Simonyan and Andrew Zisserman. Very deep convolutional networks for large-scale image recognization. In ICLR, 2015.

[23] Nicolas Vasilache, Jeff Johnson, Michael Mathieu, Soumith Chintala, Serkan Piantino and Yann LeCun. Fast convolutional nets with fbfft: a GPU performance evaluation. CoRR, abs/142.7580, 2014.

[24] Yi Yang, Ping Xiang, Jingfei Kong and Huiyang Zhou. A GPGPU compiler for memory optimization and parallelism management. In PLDI, 2010.

[25] Mattew D. Zeiler and Rob Fergus. Visualizing and Understanding Convolutional Networks. In ECCV, 2014.

[26] Chen Zhang, Peng Li, Guangyu Sun, Yijin Guan, Bingjun Xiao and Jason Cong. Optimizing fpga-based accelerator design for deep convolutional neural networks. In FPGA, 2015.

[27] NVIDIA CUDA Basic Linear Algebra Subroutines (cuBLAS) library, https://developer.nvidia.com/cublas, 2013.

[28] Kepler GK110 whitepaper: NVIDIA next generation CUDA compute architecture, 2012.

[29] Convolutional Neural Nentwork: https://en.wikipedia.org/wiki/Convolutional_neural_network

[30] NVIDIA CUDA 6.5 SDK Samples, NVIDIA, 2014

[31] Convnet-benchmarks. https://github.com/soumith/convnet-benchmarks.

[32] Tensor Processing Unit. https://en.wikipedia.org/wiki/Tensor_processing_unit

[33] NVIDIA Tesla P100 Whitepaper, 2016.